\documentclass[%
 reprint,
showpacs,
 amsmath,amssymb,
 aps,
floatfix,
nofootinbib,
]{revtex4-1}

\usepackage{graphicx}
\usepackage[caption=false]{subfig}
\usepackage{dcolumn}
\usepackage{bm}
\usepackage{hyperref}

\usepackage{amssymb,epsfig}
\usepackage{amsmath}
\usepackage{epsfig}
\usepackage{color}
\usepackage{textcomp}
\usepackage[utf8]{inputenc}
\usepackage{natbib}
\usepackage[english]{babel}
\usepackage{commath}
\makeatletter

\newcommand{\HI}{{\rm HI}}
\newcommand{\s}{{\rm s}}
\newcommand{\g}{{\rm g}}

\usepackage{hyperref}
\hypersetup{
    colorlinks=true,
    linkcolor=blue,
    filecolor=magenta,      
    urlcolor=cyan,
}

\DeclareRobustCommand{\urlfootnote}{\hyper@normalise\urlfootnote@}
\makeatother 

\begin{document}


\title{Cosmic recombination history in light of EDGES measurements of the cosmic dawn 21-cm signal  }
\author{Kanan K. Datta}
\email{kanan.physics@presiuniv.ac.in}
\author{Aritra Kundu}%
\email{aritra8354@gmail.com}
\author{Ankit Paul}%
\author{Ankita Bera}%

\affiliation{%
 Department of Physics, Presidency University, 86/1 College Street, Kolkata-700073, India
}%

\date{\today}

\begin{abstract}
The recent EDGES measurements of the global 21-cm signal from the cosmic dawn suggest that the kinetic temperature of the inter-galactic medium (IGM) might be significantly lower compared to its expected value. The colder IGM directly affects the hydrogen recombination of the universe during the cosmic dawn and dark ages by enhancing the rate of recombinations. Here, we study and quantify, the impact of the colder IGM scenario on the recombination history of the universe in the context of DM-baryonic interaction model which is widely used to explain the depth of the EDGES 21-cm signal. We find that, in general, the hydrogen ionisation fraction gets suppressed during the dark ages and cosmic dawn and the suppression gradually increases at lower redshifts until X-ray heating turns on. However, accurate estimation of the ionisation fraction requires knowledge of the entire thermal history of the IGM, from the epoch of thermal decoupling of hydrogen gas and the CMBR to the cosmic dawn. It is possible that two separate scenarios which predict very similar HI differential temperature during the cosmic dawn and are consistent with the EDGES 21-cm signal might have very different IGM temperature during the dark ages. The evolutions of the ionisation fraction in these two scenarios are quite different. This prohibits us to accurately calculate the ionisation fraction during the cosmic dawn using the EDGES 21-cm signal alone. We find that the changes in the ionisation fraction w.r.t the standard scenario at redshift $z \sim 17 $ could be anything between  $\sim 0 \%$ to $\sim 36 \%$. This uncertainty  may be reduced if measurements of HI 21-cm differential temperature at multiple redshifts are simultaneously used.  

\end{abstract}

\pacs{Valid PACS appear here}

\maketitle


\section{Introduction}
Recently, the EDGES has claimed a detection of the global HI 21-cm signal from the cosmic dawn \cite{2018bowman}. The measured signal, if confirmed independently, is found to be around two times deeper compared to the predicted signal from the standard model of cosmology and astrophysical processes. Several fundamentally different explanations have been proposed in order to explain this unusually deep signal \cite{2018barkana, 2018munoz, 2018feng, 2018pospelov, 2018mohanty}. There are also concerns about modelling the foregrounds and unaccounted systematics which can lead to significantly different interpretations  of the EDGES measurements \cite{2018hills, 2019singh}. However, a great deal of efforts have been put up to understand the 21-cm signal reported by the EDGES.  One of the most explored explanations consider interaction between the cold dark matter and baryonic gas (hereafter DM-b interaction) \cite{2018barkana, 2018munoz} which allows the heat energy to flow from the relatively warm baryonic gas to the colder dark matter. Consequently,  the baryonic gas cools much faster compared to the standard adiabatic cooling rate \cite{2015munoz} and could explain the depth of the EDGES signal. This possibility opens up a new and promising avenue to constrain the DM properties such as  the mass of dark matter particles and the interaction cross section. A large number of studies focus on this aspect and put tight constraints on the DM properties \cite{2018slatyer, 2018fialkov, 2018damico, 2018fraser}.

The `colder inter galactic medium (IGM)' scenario has direct impact on the hydrogen cosmic recombination history of the universe. The hydrogen recombinations in the universe start very early at redshift $z \sim 1500$ and the universe becomes largely neutral by redshift $z \sim 1000$. However, a very small fraction ($\sim 5\%$) of hydrogen still remains in ionised form at redshift $z \sim 1000$ and it gradually becomes neutral over the redshift range $1000 > z \gtrsim 20$. The residual electron plays an important role in determining the IGM temperature and the HI differential brightness temperature during the dark ages and cosmic dawn.  The hydrogen recombination rate which determines the residual electron fraction (or the hydrogen ionisation fraction) increases in the colder IGM scenario. This results in faster recombination and lower ionisation fraction. 

Detailed understanding of the cosmic recombination history is of immense importance and a lot of efforts, since late 60s, has been invested for its accurate calculations \cite{1968Peebles, 1968zeldovich, 1985jones, 1990krolik,1999Seager, 2000Seager}. State of the art codes have been developed to keep the errors in estimating the hydrogen ionisation fraction at $\sim 1 \%$ level \cite{recfast, hyrec}. Accurate knowledge of the ionisation fraction during the cosmic dawn and dark ages is crucial for understanding the roles of the cosmic magnetic field on the thermal history of the IGM \cite{2005sethi, bera2020} and the large scale structure formation \cite{2009sethi}, formation of molecules in the early universe \cite{2002lepp}, contributions of ionisation fluctuations to the total fluctuations in HI 21-cm field \cite{2018ansar}. 

In this paper, we explore the impact that the colder IGM scenario has on the recombination history of the universe, particularly during the cosmic dawn and dark ages. We study this in the context of the DM-b interaction model introduced in \cite{2014tashiro, 2015munoz}. It is shown that  interactions between the DM and baryons along with the standard heating and cooling processes can explain the depth of the EDGES 21-cm profile \cite{2018barkana, 2018munoz}. We calculate the thermal history of the IGM for all possible combinations of the DM-b interaction model parameters (mass of the dark matter particle and the interaction cross section) and compare with the EDGES 21-cm signal. We then focus on the changes in the hydrogen ionisation fraction from the standard scenario, i.e., without the interaction, for model parameters which are consistent with the EDGES 21-cm signal. Next, we discuss the prospects of more precise measurements of the cosmic dawn 21-cm signal in the accurate estimation of the ionisation fraction. We also discuss roles of various physical processes that determines the thermal and ionisation history during the cosmic dawn and dark ages. 

The paper is organised as  follows. In the next two sections, we set up the basic equations for calculating the global 21-cm signal and briefly discuss the EDGES measurements respectively. Next, we present the equations for calculating the evolution of the IGM, dark matter  temperature and hydrogen ionisation fraction in the context of DM-b interaction.  We present our results on the thermal history and study the impact of the colder IGM on the ionisation history in the subsequent section. Finally, we summarise and discuss our results. We use cosmological parameters $\Omega_{\rm m0}=0.3$, $\Omega_{\Lambda}=0.7$, $\Omega_{\rm b0}=0.0486$, $h=0.667$ which are consistent with measurements by the PLANCK experiment \cite{planck15}. 
\label{sec:intro}

\section{Global HI 21-cm signal}
\label{sec:sec2}
The global  HI 21-cm  differential brightness temperature   at  redshift $z$ can be written as \citep{2006Furlanetto}
\begin{equation} \label{tb}
\begin{aligned}
\frac{T_{b}}{\rm mK} \approx 27 x_{\HI} \left( \frac{\Omega_{b0}h^2}{0.023} \right) \left( \frac{0.15}{\Omega_{m0}h^2 }\frac{1+z}{10} \right) ^{0.5}\left( \frac{T_{\s}-T_{\gamma}}{T_{s}} \right) ,
\end{aligned}
\end{equation}
where $T_{\gamma}$  and $x_{\HI}$ are the cosmic microwave background radiation (CMBR) temperature and hydrogen neutral fraction respectively.  The spin temperature $T_{s}$ is a measure of the ratio of ground state HI atoms in the triplet and singlet states. It is determined by three physical processes, 1. radiative coupling with the background CMBR, 2.  coupling with the gas kinetic temperature $T_{\g}$ through collisions and, 3. coupling with $T_{\g}$ through  Ly-$\alpha$ photons (also known as Wouthuysen - Field coupling). Therefore,  the spin temperature, in general, can be  represented as\citep{1958Field}
\begin{equation} \label{ts}
T_{\s}^{-1} = \left[\frac{(x_{\rm c}+x_{\alpha})T_{\g}^{-1}+T_{\gamma}^{-1}}{1+x_{\rm c} +x_{\alpha} }\right],
\end{equation}
where $x_{\rm c}$ and $x_{\alpha}$  are the collisional and Ly$\alpha$ coupling coefficients respectively \citep{2006Furlanetto}.  In the standard picture the collisional coupling starts to dominate over the other two processes at redshifts $z \sim 200$. During the cosmic dawn, the first generation of star starts to emit radiation including the Ly-$\alpha$ photons.  This makes the Ly-$\alpha$ coupling strong and, as a consequence, the spin temperature is expected to  be coupled to the gas kinetic temperature, i.e., $T_{\s} \approx T_{\g} $ at redshift $z \lesssim 20$. 

\section{EDGES measurements and colder IGM}The EDGES experiment has claimed to have detected the global HI 21-cm differential brightness temperature $T_{b}$  in the redshift range of $15 \lesssim z \lesssim 22$ \cite{2018bowman}. The measured signal is found to match a flattened gaussian function centred around redshift $z=17.2$ and has a best fit amplitude of $-500 \, {\rm mK}$. It is, therefore,  likely that the Ly-$\alpha$ coupling becomes very strong, i.e., $T_{\s} \approx T_{\g} $ by redshift $z \sim 17$. The experiment also reported that  the amplitude of the signal should be between $-300 \, {\rm mK}$ and $-1000 \, {\rm mK} $ (with $99 \%$ confidence) if the uncertainties due to thermal and systematic noise are considered. One possible explanation could be that the IGM temperature at redshift $z = 17.2$ should  lie between $1.76 \, {\rm K}$  and $ 5.4$ K (with $99 \%$ confidence).  However, according to the  known standard cosmology and astrophysical processes  the IGM temperature at redshift $z=17.2$ should be around $7$ K  which is significantly higher than that found by the EDGES. The excess radio background measured by ARCADE  2 \cite{fixsen2011}can also account for larger than normal amplitude of the signal \cite{2018feng}.

\section{IGM temperature and recombination history}
In order to calculate the accurate recombination history, we need to know the full thermal history of the IGM.  The evolution of the IGM kinetic temperature, in the backdrop of the DM-b interaction, can be calculated using following equation \cite{2015munoz},. 
\begin{align}
    \frac{dT_g}{dz} &= \frac{2 T_g}{1 + z} - \frac{32 \sigma_T \sigma_{SB} T^4_0}{3 m_e c^2 H_0 \sqrt{\Omega_{m0}}} (T_{\gamma} - T_g)(1 + z)^{3/2} \frac{x}{1 + x} \nonumber \\
      &\qquad {} - \frac{2}{3 k_B n_H}\frac{f_{\rm heat}\epsilon_X}{H(z)(1+z)} - \frac{2}{3 k_B}\frac{\dot{Q_b}}{H(z)(1+z)} 
    \label{eq:Tg}      
\end{align}

The first and second term in the rhs correspond to the cooling due to adiabatic expansion of the universe and heating due to heat flow from the CMBR through its interaction with free electrons. The third and fourth term correspond to the heating due to X-ray photons \cite{pritchard2007} and  cooling/heating due to interactions between the baryonic IGM and dark matter respectively.  $\sigma_{\rm T}$,  $\sigma_{\rm SB}$  are the Thomson scattering cross-section,  Stefan Boltzmann constant respectively. $T_0$ is the CMBR temperature at present. $\epsilon_X$ is the total rate of energy deposited in the IGM  per unit volume due to X-ray photons. $f_{\rm heat}$ is the fraction of the total deposited energy used for IGM heating. ${\dot{Q_{\rm b}}} $ is  the heating/cooling rate of baryons. The evolution of the dark matter temperature $T_{\chi}$ is similar to the above except the fact that the dark matter does not interact with the CMBR.  Hence we can write
\begin{equation}
\frac{dT_\chi}{dz} = \frac{2 T_\chi}{1+z} - \frac{2}{3 k_B}\frac{\dot{Q_\chi}}{H(z)(1+z)},
\label{eq:Tx}
\end{equation}
where ${\dot{Q_{\chi}}} $ is  the heating rate of dark matter. We model both the quantities ${\dot{Q_{\rm b}}} $ and ${\dot{Q_{\chi}}}$ as prescribed in \cite{2015munoz} ( see equation \ref{eq:heatrate} and subsequent texts of this paper for details). The evolution of the ionisation fraction $x$ can be calculated using the equation \cite{1968Peebles}
\begin{align}
\frac{dx}{dz} &= \frac{C}{H(z)(1 + z)} \Big[ \alpha_e (T_{\g}) x^2 n_H  \nonumber  \\
      &\qquad {} - \beta_e (T_{\g}) (1 - x) e^\frac{-h_p \nu_\alpha}{k_B T_g} \Big] -\frac{ f_{\rm ion} \epsilon_X}{H(z)(1 + z) n_H E_{\rm th}}.
\label{eq:ion}      
\end{align}
The first and second terms inside the third brackets quantify hydrogen recombination and photoionization due to CMBR respectively.  The last term accounts for ionisation in the IGM due to X-ray photons.  We assume that X-ray photons are coming from the first generation starburst galaxies. The ionisation fraction is defined as  $x=n_{e}/n_{\rm H}$. Here, we assume $n_{\rm H} = n_{\rm HI}+ n_{\rm HII}$ and $n_{\rm HII} = n_e$, assuming helium to be fully neutral for the redshift range of our interest. $C = \frac{1+ K \Lambda (1-x) n_{H}}{1+K(\Lambda+\beta_{e})(1-x) n_{H} }$ is the probability of a  hydrogen atom jumping to the ground state from the first excited state without exciting an adjacent ground state atom, $\Lambda=8.3 \, {\rm s}^{-1}$ is  the transition rate from 2s $\rightarrow$ 1s  through the two photon decay.  Redshifting of Ly-$\alpha$ photons due to universe's expansion  is accounted by $K=\frac{\lambda_{\alpha}^{3}}{8\pi H(z)}$.  ${h_{p}  \nu_{\alpha}=10.2 \, {\rm eV}}$, $\alpha_{e}$ and $\beta_{e}$ are  the hydrogen recombination and photo-ionisation coefficients respectively. We note that both $\alpha_{e}$ and $\beta_{e}$ depend on the IGM temperature $T_{\rm g}$.  We use $\alpha_e (T_{\g})=F \times 10^{-19} \frac{a t^b}{1 + c t^d} \hspace{0.1cm} m^3 s^{-1} $, where $a = 4.309$, $b = -0.6166$, $c = 0.6703$, $d = 0.53$, $F = 1.14$ is the fudge factor  and $t = \frac{T_g}{10^4} \, {\rm K}$ \cite{1999Seager, 2000Seager}. We calculate the photoionisation coefficient  $\beta_e$ from $\alpha_e$ using the relation $\beta_e (T_{\g}) = \alpha_e \Bigg( \frac{2 \pi m_e k_B T_g}{h^2_p} \Bigg)^{3/2} e^{-E_{2s}/k_B T_g}$. Ionisation rate per unit volume due to X-ray photons is approximately obtained by multiplying a factor $f_{\rm ion} /E_{\rm th} $ to the quantity $\epsilon_X$ \cite{pritchard2007}. We use the fitting formulae provided in \cite{shull1985} to calculate the fraction of X-ray energy that goes into heating $f_{heat}$ and ionisation $f_{ion}$. Here, we ignore ionisation due to UV photons. UV photons from first galaxies also ionize surrounding HI. However, due to their shorter mean free paths those ionized regions  will remain confined very close to sources during cosmic dawn. It is very unlikely that UV photons will cause significant changes to the ionisation state in the bulk IGM during the cosmic dawn \cite{pritchard2007}. On the contrary, X-ray photons penetrate very deep in the IGM due to their large mean free paths and affect both thermal and ionisation state of the IGM.

We follow the formalism presented in \cite{pritchard2007} for studying the impact of X-ray photons. We calculate the total rate of X-ray energy deposited in the IGM as
\begin{equation}
\epsilon_X=4\pi n_i \int d\nu\sigma_{\nu,i}(h\nu-h\nu_{th}) J_X(\nu, z),
\label{eq:epsilonX}
\end{equation}
where the index `i' denotes various species such as HI, HeI, HeII, $n_i$ is the number density of the species i, $\sigma_{\nu,i}$ is the corresponding photo-ionisation cross section,  $\nu_{th}$ = $E_{th}/h$ is the threshold frequency for photo ionisation. The flux of X-ray photons per unit frequency at frequency $\nu$ and redshift $z$ can be written as
\begin{flalign}
J_X(\nu, z) = \int_{z}^{z'} dz' \frac{(1+z)^2}{4\pi} \frac{c}{H(z')} \hat\epsilon_X(\nu', z')e^{-\tau}
\label{eq:Jx}
\end{flalign}
where $\hat\epsilon_X(\nu', z')$ is the comoving photon emissivity for X-ray sources, and $\nu'$ is the emission frequency at $z'$ which is redshifted to frequency $\nu$ at $z$, i.e., $\nu'=\nu\frac{(1+z')}{(1+z)}$. The optical depth is given by,
\begin{align}
\tau(\nu,z,z') &= \int_{z}^{z'}\frac{dl}{dz''}dz''[n_{HI}\sigma_{HI}(\nu'') + n_{HeI}\sigma_{HeI}(\nu'')]\nonumber \\
    &\qquad {} \hspace{1.5 cm}
\label{eq:tau}
\end{align}
We consider starburst galaxies as the source of X-ray photons. We link the source  emissivity per unit comoving volume per unit frequency to the star formation rate density (SFRD) as
\begin{equation}
\hat\epsilon_X(\nu,z)=\hat\epsilon_X(\nu)\Bigg(\frac{\rm SFRD}{M_{\odot} \hspace{0.1cm} yr^{-1} \hspace{0.1cm} Mpc^{-3}}\Bigg).
\end{equation}
The spectral distribution function is modelled as,
\begin{equation}
\hat\epsilon_X(\nu)= \frac{L_0}{h\nu_0}\Bigg(\frac{\nu}{\nu_0}\Bigg)^{-\alpha_S - 1},
\end{equation}
where $\alpha_S$ is the power law index, $h\nu_0$ = 1 keV. We set $L_0$ = 3.4 x 10$^{40}$ $f_X$ erg s $^{-1}$ Mpc$^{-3}$ and choose $\alpha_S=1.5$ for starburst galaxies \cite{rephaeli1991, oh2001, furlanetto2006}. We also set $f_x=1$ which keeps the total X-ray luminosity per unit star formation rate consistent with observations of present day starburst galaxies. The star formation rate density is modelled as,
\begin{equation}
{\rm SFRD} = \Bar{\rho}^{0}_{b}(z)f_{\star}\frac{df_{coll}(z)}{dt},
\end{equation}
where $\Bar{\rho}^{0}_{b}$ and $f_{\star}$ are the present day mean baryon density and start formation efficiency respectively.

\section{DM-baryon interaction} We discussed in the previous section that, in order to explain the EDGES 21-cm profile, the IGM temperature needs to be lowered by a factor of $\sim 2$ at redshift $z \sim 17$ which can be achieved by considering interactions between the cold dark matter and baryons. However, the hydrogen recombination rate gets enhanced for colder IGM  and this results in relatively higher neutral fraction, i.e., lower ionisation fraction. As a consequence the heat transfer rate from the CMBR to baryon gets reduced. We need to solve all these coupled effects simultaneously.   

We now focus on our modelling of the DM-baryonic interaction which we follow from \cite{2015munoz}. The cooling/heating rate of baryons due to DM-baryonic interactions are calculated as follows
\begin{align}
\frac{dQ_b}{dt} &= \frac{2 m_b \rho_{\chi} \sigma_0 e^{-r^{2} / 2} (T_{\chi} - T_g) k_B c^4}{(m_b + m_{\chi})^2 \sqrt{2\pi} u^{3}_{th}} \nonumber \\
&\qquad {} + \frac{\rho_{\chi}}{\rho_m} \frac{m_{\chi} m_b}{m_{\chi} + m_b} V_{\chi b} \frac{D(V_{\chi b})}{c^2}.
\label{eq:heatrate}
\end{align}
The equation governing the heating rate of the dark matter $\dot{Q_{\chi}}$ can be obtained by flipping $\chi \leftrightarrow b$ in the above equation. The interaction cross-section has been parameterised  as $\sigma = \sigma_0 (v/c)^{-4}$. This kind of scaling of the cross section with the velocity $v$ is very effective in transferring the heat energy from baryonic gas to the dark matter without significantly affecting other episodes of the cosmic evolution. The first term in the rhs makes sure that heat energy flows from the warmer fluid (in our case the baryon) to the colder fluid (the dark matter).  We see that the heating rate is proportional to ($T_{\chi}-T_{\rm g}$), i.e., the temperature difference between the two fluids.  This will try to make the temperatures of the two fluids equal. The second term accounts for the heating caused due to friction between the dark matter and baryonic fluids. The dark matter and baryonic fluid flow at two different velocities which produces friction between the two and that heats up both the fluids. Note that this heating depends on the relative velocity between the two fluids $V_{\chi b}$ and drag $D(V_{\chi b} )$ which can be calculated using the equations

\begin{equation}
\frac{dV_{\chi b}}{dz} = \frac{V_{\chi b}}{1+z} + \frac{D(V_{\chi b})}{H(z)(1+z)}
\label{eq:vxb}
\end{equation}
and
\begin{equation}
D(V_{\chi b})  = \frac{\rho_m \sigma_0 c^4}{m_b + m_\chi} \frac{1}{V^2_{\chi b}} F(r),
\end{equation}

where $\rho_{\chi}$, $\rho_m$ are the energy densities of the dark matter and the total matter respectively, $m_b$ and $m_\chi$  are the masses of the baryonic and dark matter particles respectively. `$r$' is defined as $r = \frac{V_{\chi b}}{u_{th}}$, where $u_{th} = c \sqrt{k_B(T_b/m_b + T_\chi/m_\chi)}$. The function $F(r)$ is defined as

\begin{equation}
F(r) = erf \Big( \frac{r}{\sqrt{2}} \Big) - \sqrt{ \frac{2}{\pi}} r e^{-r^2/2}.
\end{equation}

$F(0) = 0$ and $F(r) \rightarrow 1$ when $r \rightarrow \infty$.

We see from eq. \ref{eq:Tg} and \ref{eq:heatrate}  that the IGM temperature $T_{\g}$ depends on the relative velocity $V_{\chi b}$. As a consequence the global  HI 21-cm  differential brightness temperature $T_{b}$ and the ionisation fraction $x$ also become $V_{\chi b}$ dependent. The initial value $V_{\chi b, 0}$ follows the probability distribution $P(V_{\chi b, 0}) = \frac{e^{-3 V^2_{\chi b, 0}/( 2 V^2_{\rm rms})}}{(2 \pi V^2_{\rm rms}/3)^{3/2}}$. We calculate the velocity averaged differential brightness temperature  using

\begin{equation}
\langle T_b (z) \rangle = \int d^3 V_{\chi b} T_b (V_{\chi b}) P(V_{\chi b}).
\end{equation}

Similarly, we calculate the velocity averaged IGM temperature $\langle T_g (z) \rangle$ and ionisation fraction  $\langle x \rangle$. $V_{\rm rms}$ is assumed to be $29 \, {\rm km/s}$ at the initial redshift $z=1010$ \cite{2014alihaimoud}.

 \section{Results}
 We solve eqs. \ref{eq:Tg}, \ref{eq:Tx}, \ref{eq:ion} and \ref{eq:vxb} simultaneously after setting the initial conditions at redshift  $z=1010$.  At the starting redshift $z=1010$, we assume $T_{\g}=T_{\gamma}$ and $T_{\chi}=0$. We obtain the initial ionisation fraction $x(z=1010)=0.055$ from the RECFAST code \cite{recfast}. As discussed above the initial  relative velocity $V_{\chi b, 0}$ follows a Gaussian probability distribution with the rms of $\sim 29 \, {\rm km/s}$ and we average over the relative velocity to obtain various temperatures and the ionisation fraction. 
 \begin{figure}[hbtp]

    \subfloat[]{%
     \includegraphics[width=9.0cm,angle=0]{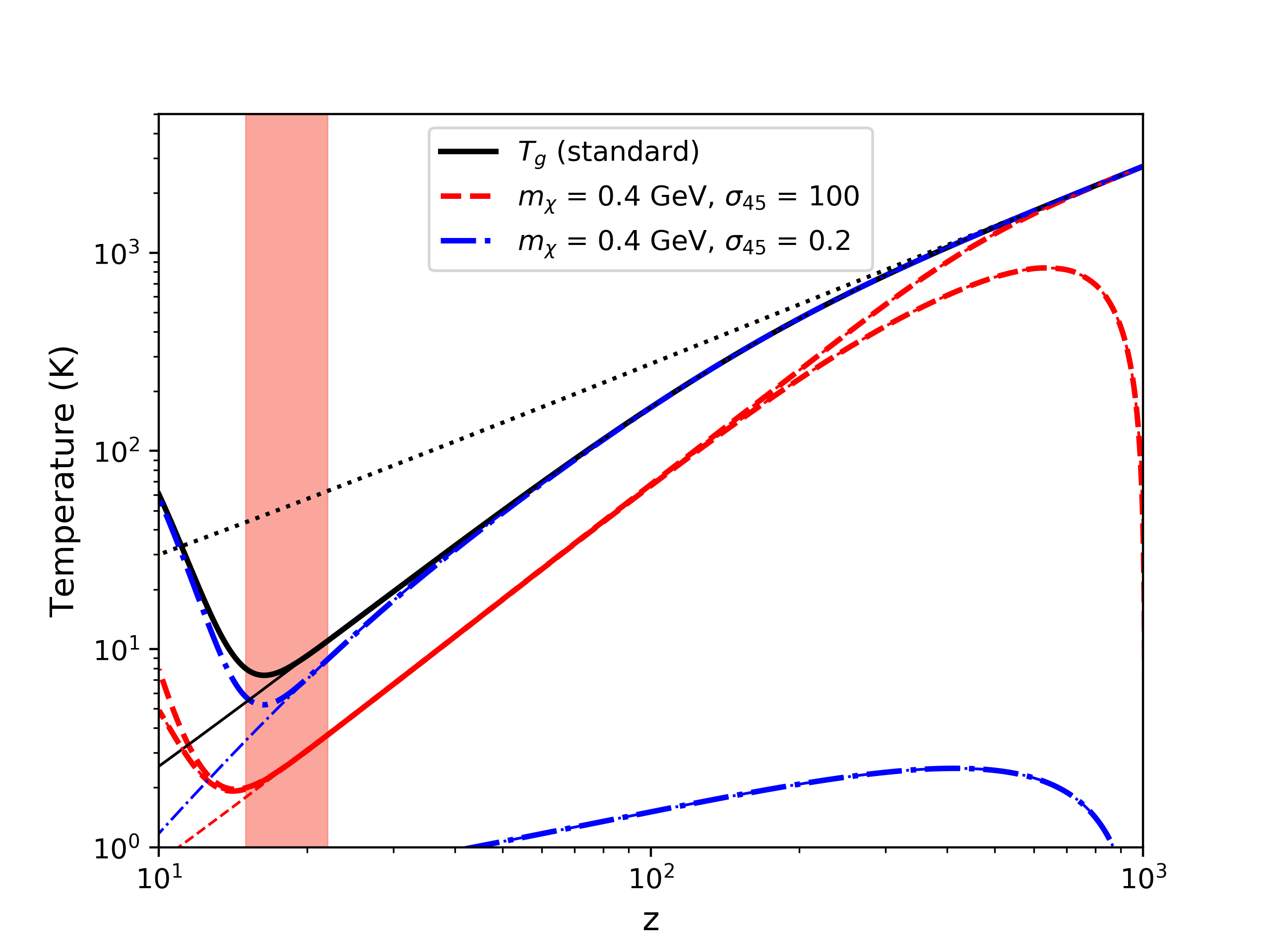}%
    }

    \subfloat[]{%
    \includegraphics[width=9.0cm,angle=0]{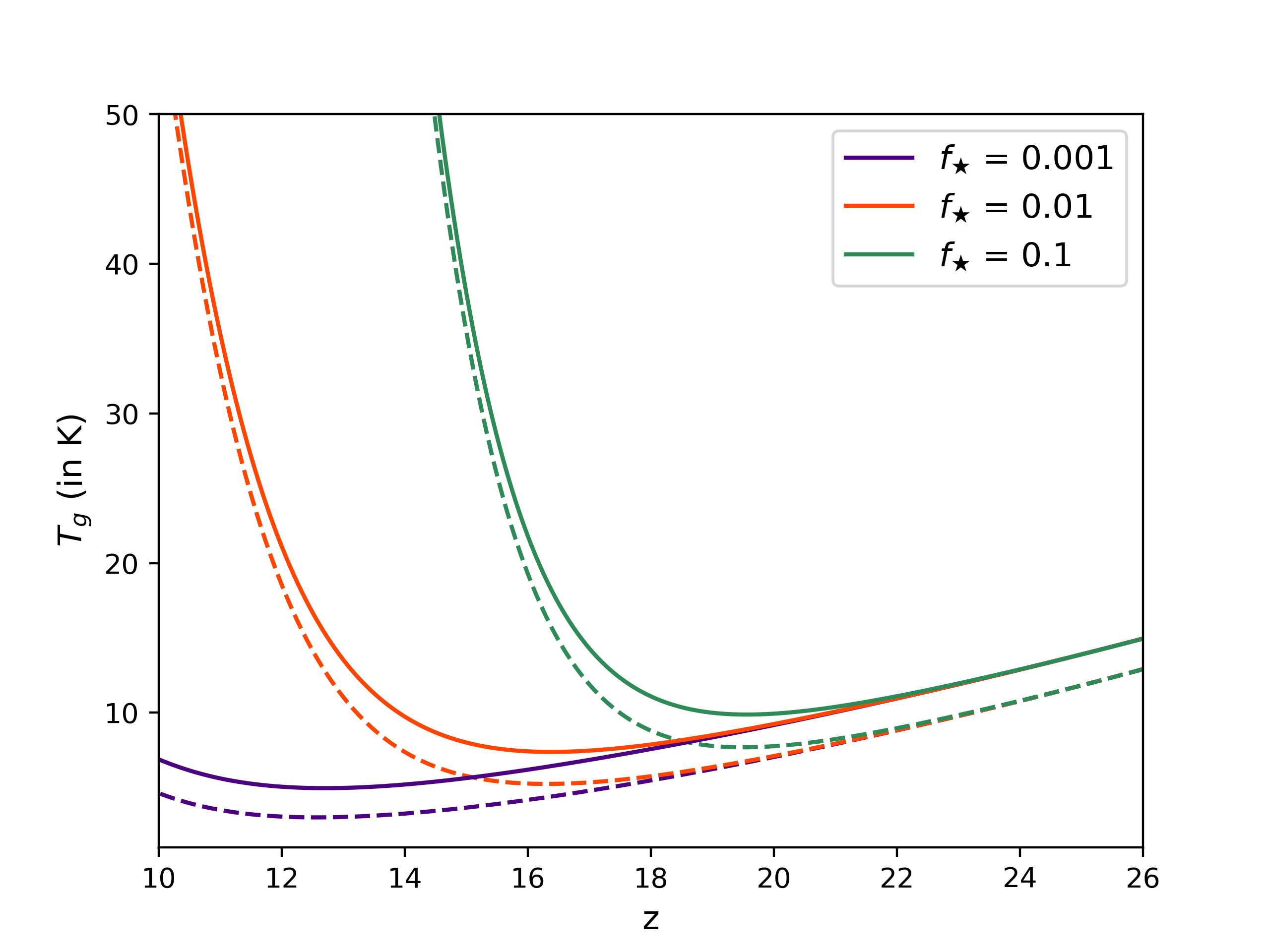}%
    }
    \caption{The upper panel (a) shows the evolution of IGM temperature, dark matter temperature for scenarios where the DM-b interactions are included and compares with the standard scenario without the  DM-b interaction (thin black-solid line). The thick and thin lines show results with and without X-ray heating respectively. The red-dashed lines and blue-dash-dotted lines correspond to $(m_{\chi}/{\rm GeV}\, ,\sigma_{45}) =(0.4, 100)$ and $(0.4, 0.2)$ respectively. We use $f_{\star}=0.01$ for calculating the X-ray heating. The upper and lower lines (both thick and thin) in each set represent the IGM and dark matter temperature respectively. The HI differential brightness temperature $T_b$ predicted using these two sets of parameters are consistent with the EDGES 21-cm profile measurements at redshift $z=17.2$. The black-dotted line shows the CMBR temperature. The shaded region shows the redshift range covered by the EDGES measurements. The lower panel(b) plots the evolution of the IGM temperature for three different values of star formation efficiency $f_{\star}$. The solid lines correspond  to the standard scenario and the dashed lines correspond to a colder IGM scenario with $m_{\chi}=0.4$ and $\sigma_{45}$ = 0.2.}
    \label{fig:temp}
 \end{figure}
 
\subsection{Evolution of IGM temperature} 
First, we discuss our results on how the DM-b interaction and X-ray photons from first sources alter the evolution of the IGM temperature from the standard predictions. This helps us to understand the impact these effects have on the recombination history. Fig. \ref{fig:temp}a shows the evolution of the IGM and the dark matter temperature. Thin dashed and dot-dashed curves correspond to ($m_{\chi}$, $\sigma_{45}$)= $(0.4\, {\rm GeV}, 100)$ and $(0.4 \, {\rm GeV}, 0.2)$ respectively, where $\sigma_{45}=\frac{\sigma_0}{10^{-45} \, {\rm m^2}}$. These do not include effects of X-ray photons. Fig. \ref{fig:temp}a also shows the IGM temperature ( thin solid black lines) for the standard scenario  which does not include neither the DM-b interaction nor the X-ray heating. Results for X-ray heating are shown in thick curves. The three competitive processes, i.e.,  interaction between baryons and the CMBR, X-ray heating and interaction between baryons and the DM, together determine the resultant IGM temperature. The interaction between the CMBR and baryons tries to keep the IGM temperature the same as the CMBR temperature. On the other hand, the DM-b interaction tries to bring the dark matter and IGM temperatures to some other thermal equilibrium with temperature $T_{\rm eq}$ which is lower than the CMBR temperature. For a large value of the DM-b cross-section $\sigma_{45}$, the DM and baryon reach to a thermal equilibrium much faster at higher redshift. We see in Fig. \ref{fig:temp}a that the thermal equilibrium between the dark matter and baryon is reached by redshift $z \sim 200$ for $\sigma_{45}=100$. The large value of $\sigma_{45}$ also helps the IGM to decouple from the CMBR earlier at redshift around $z \sim 500$. After attaining the equilibrium the dark matter and baryonic gas remain thermally coupled for the rest of the redshift range we explore and the equilibrium temperature scales as $(1+z)^2$. The early decoupling of the IGM temperature from the CMBR  helps it to cool faster and explain the EDGES results. For smaller cross section, the interaction between the DM and baryons tries to bring the DM and the IGM temperature to equilibirum at a later redshift. Here, the CMBR-baryon interaction dominates over the DM-b interaction and the evolution of IGM temperature follows the standard model upto very late. However, at later times the CMBR-baryon interaction becomes weaker and the DM-b interaction becomes dominant. Consequently, the IGM temperature decouples from the CMBR temperature and both the IGM and dark matter temperature approach to each other. In this phase the IGM temperature falls very rapidly as redshift decreases which we see from Fig. \ref{fig:temp}a, from curves corresponding to $\sigma_{45}=0.2$ and $m_{\chi}=0.4 \, {\rm GeV}$. Although, in this case, the IGM temperature follows the standard prediction up to redshift $z \sim 40$, it drops after that. We note that this parameter set too predicts colder IGM and explain the depth of the EDGES 21-cm signal measurements at redshifts $z \sim 17$ . However, we see that the thermal history of the IGM according to the later parameter set is quite different from the first.  The first one predicts colder IGM for a longer period of the cosmic time whereas the IGM is colder for a shorter period in the second case. The `drag heating' term heats up both the IGM and the dark matter irrespective of their individual temperatures. We notice that this helps the IGM temperature to get coupled with the dark matter temperature earlier.  

The above picture changes after X-ray photons from the first stars begin to interact with the baryons. Thick  curves in Fig. \ref{fig:temp}a show results when the effect of X-ray heating is included. We  set  $f_{\star}=0.01$ in Fig. \ref{fig:temp}a. It is also important to note that the impact of X-ray heating on the IGM temperature remains negligible at higher redshifts $z \gtrsim 17$. This implies that evolution of ionisation fraction $x$ will remain unaffected at higher redshifts. We observe that $T_g$ starts rising at redshift $z \sim 17$. This is roughly consistent with the EDGES profile. We further note that the IGM temperature at redshift $z \lesssim 17$ for  $(m_{\chi}, \sigma_{45}) = (0.4 \, {\rm GeV}, 100)$ is significantly lower compared to the scenario with $(m_{\chi}, \sigma_{45}) = (0.4 \, {\rm GeV}, 0.2)$. Both scenarios are consistent with the EDGES findings at $z\sim 17$ but predict significantly different IGM temperature at lower redshifts. Precise measurements of IGM temperature at relatively lower redshifts can be used to further constrain dark matter baryon interaction model. Fig \ref{fig:temp}b shows the evolution of $T_g$ for three different values of $f_{\star}$ i.e., $0.001,\, 0.01$ and $0.1$. The IGM temperature starts rising as early as at redshift $z\sim 20$ and crosses the CMBR temperature at $z\sim 15 $ for $f_{\star}=0.1$, whereas for $f_{\star}=0.001$, these happen at lower redshifts. Both these cases are inconsistent with the EDGES findings. In order to keep the IGM temperature consistent with the EDGES findings, X-ray heating can not dominate at $z \gtrsim 17$. This implies that $f_{\star}$ should be $\lesssim 0.1$.

\subsection{Global 21 cm signal} 
\begin{figure}[hbtp]
\includegraphics[width=8.5cm,angle=0]{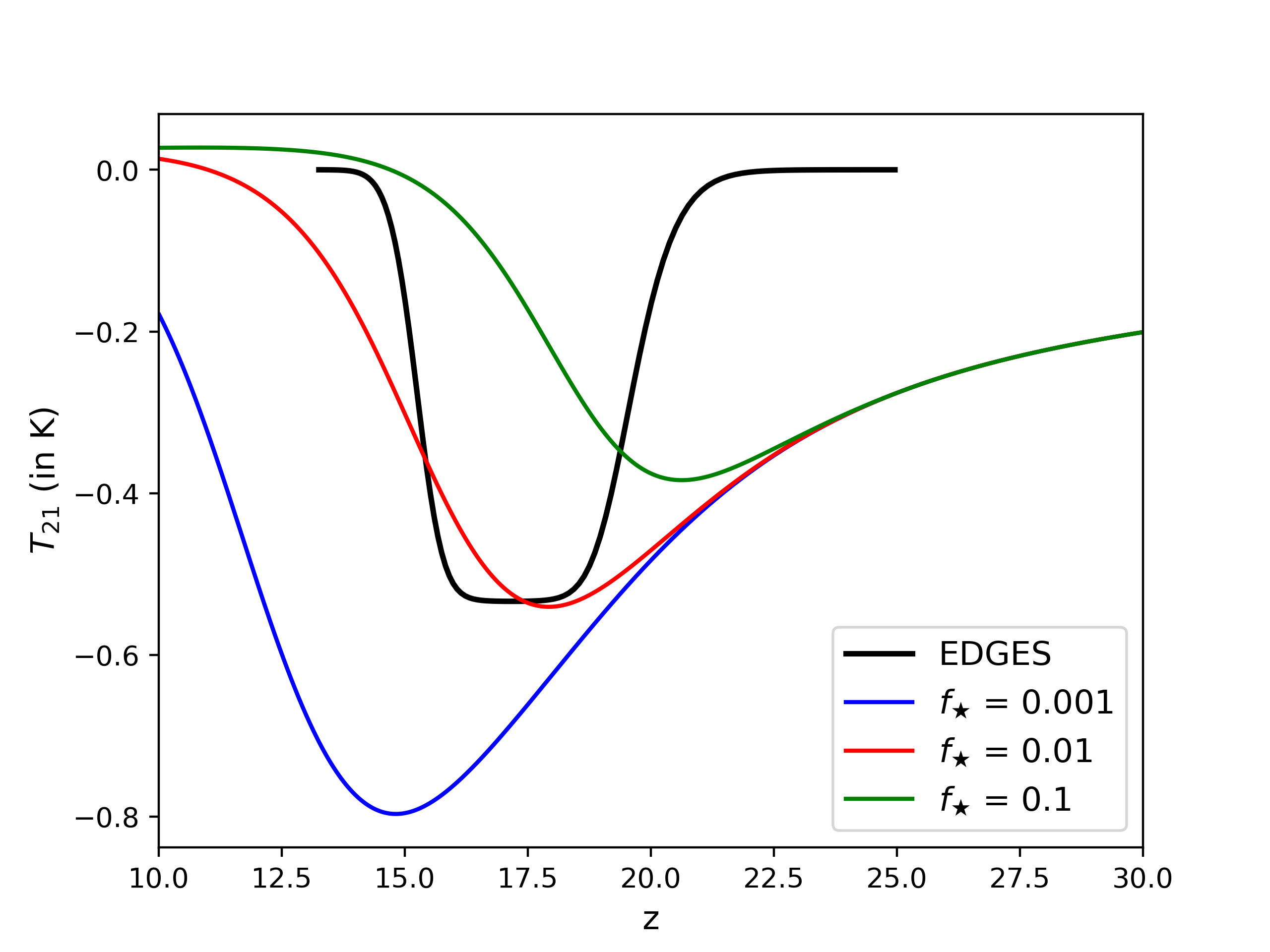}

 \caption{This plot shows the redshift evolution of the HI differential brightness temperature $T_b$ for three different values of $f_{\star}$ in the colder IGM scenario. Here we use $m_{\chi}$ = 0.4 GeV, $\sigma_{45}$ = 0.2.}
\label{fig:T_21}
\end{figure}

Fig. \ref{fig:T_21} plots the differential brightness temperature $T_b(z)$  for $f_{\star}=0.001$, $0.01$ and $0.1$ and compares those with the EDGES absorption profile. These also include the DM-baryon interaction with parameters $m_{\chi}$ = 0.4 GeV, $\sigma_{45}$ = 0.2. We see that $T_b(z)$ starts rising at redshift $z \sim 20$ for $f_{\star}$ = 0.1 which is significantly higher in compare to what EDGES has found. On the other hand $T_b(z)$ starts rising very late and the signal amplitude is very high for $f_{\star}$ = 0.001 as the X-ray heating is not efficient at earlier times. We, however, observe that the global 21-cm signal obtained from our model for $f_{\star}$ = 0.01 is consistent with the amplitude of the EDGES profile and also the redshift at which the signal rises. We further note that the model signal cannot explain the shape of the EDGES profile and further exploration is needed towards this aspect.

\begin{figure}[hbtp]
    \subfloat[]{%
    \includegraphics[width=9.0cm,angle=0]{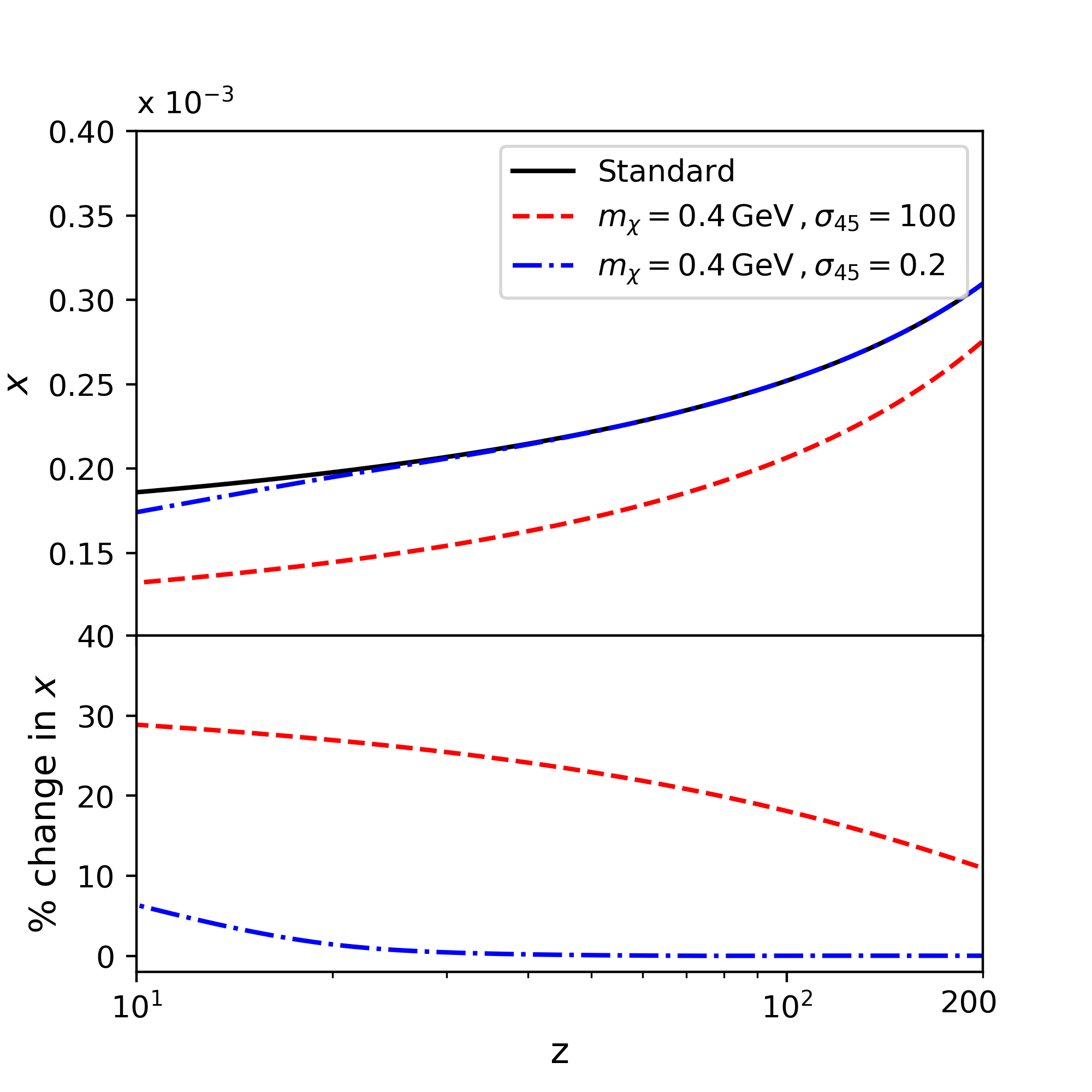}%
    }
    \vspace{0.01 cm}
    \subfloat[]{%
        \includegraphics[width=9.0cm,angle=0]{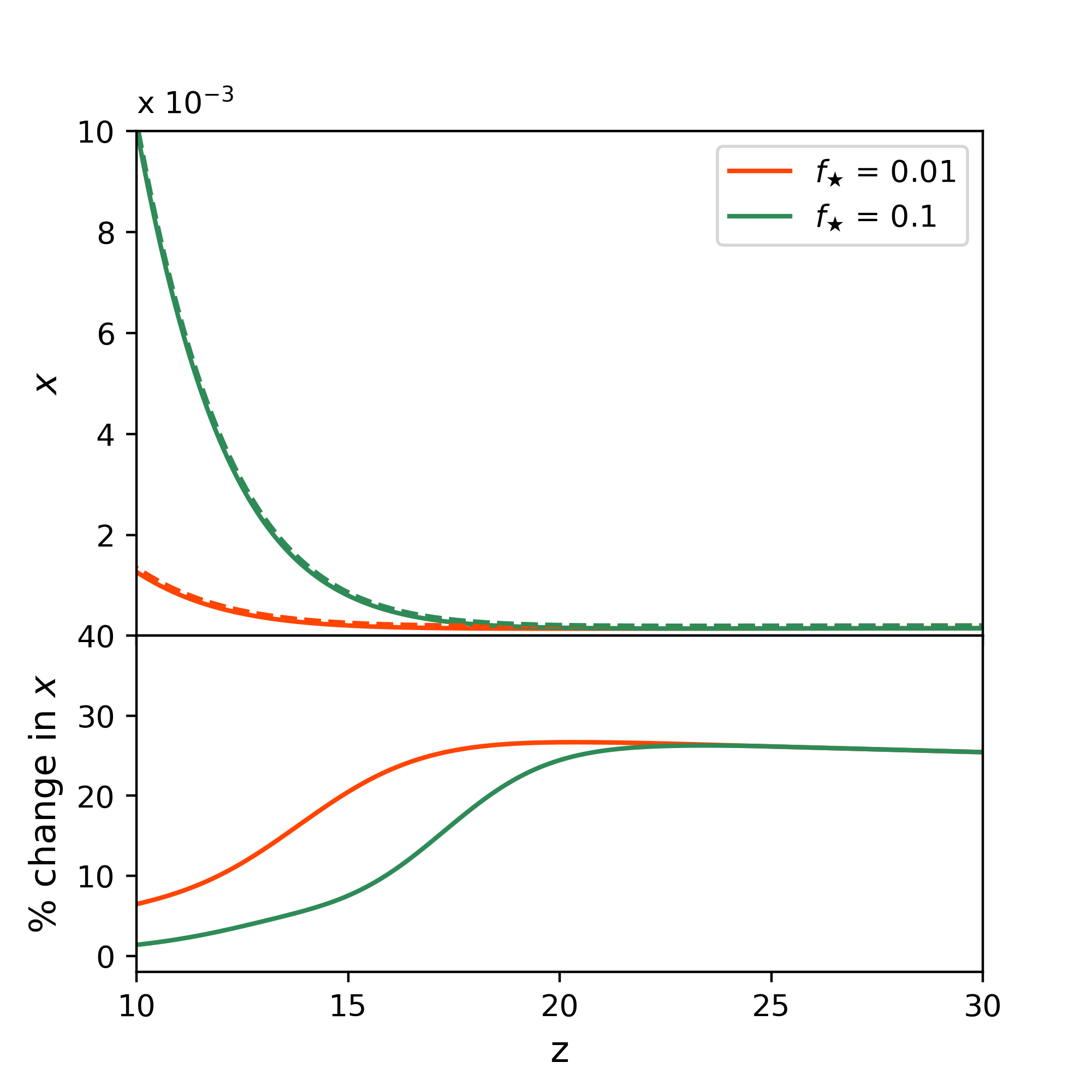}%
 }
 
    \caption{The upper panel (a) shows the evolution of ionisation fraction for the same sets of $(m_{\chi}/{\rm GeV}\, , \sigma_{45})$ parameters as in Fig. \ref{fig:temp} and compares these with the standard case i.e., no DM-b interaction and no X-ray heating scenario (black solid line). The lower panel of `a' shows the percentage difference between the above two. Fig. 3(b): The upper panel shows the ionisation fraction when the effects due to both DM-baryon interaction and  X-ray photons are included for $f_{\star} = 0.1$ and $0.01$. The lower panel shows the percentage change in the ionisation fraction w.r.t. the case when the DM-baryon interaction is excluded. The parameters used here are $m_{\chi} = 0.4 \, {\rm GeV}$ and $\sigma_{45} = 0.2$ .}
    \label{fig:xfrac}
\end{figure}

 \subsection{Recombination history} We now focus on the impact of the colder IGM on the evolution of ionisation fraction, i.e., the recombination history. The upper panel of Fig. \ref{fig:xfrac}a shows the evolution of the ionisation fraction in presence of DM-baryonic interaction with the same two sets of $m_{\chi}$ and  $\sigma_{45}$ used above. We do not include the X-ray heating here. As expected, both the parameter set, which are consistent with the EDGES 21-cm signal, predict lower ionisation fraction compared to that predicted in the standard scenario. However, we notice, in the lower panel of Fig. \ref{fig:xfrac}a, that the percentage change in the ionisation fraction $x$ at redshift $z \sim 17.2$ are  $\sim 27\%$ and $\sim 2.1 \%$ respectively although the differential brightness temperatures $T_b$ are very similar ($-726$ mK and $-668$ mK respectively) in these two scenarios. This apparent ambiguity can be explained if we look at the evolution of the IGM temperature in these two scenarios presented in Fig. \ref{fig:temp}a. The IGM temperature $T_{\g}$ for the first case ($m_{\chi}=0.4, \sigma_{45}=100$)  deviates from the standard scenario much earlier  and remains deviated for the rest of the redshift range of interest. This helps the recombination rate to remain higher for an extended period of the cosmic time and, as a result, the ionisation fraction is considerably lower. On the other hand, for the second scenario ($m_{\chi}=0.4, \sigma_{45}=0.2$) the evolution of the IGM temperature  is very similar to the standard scenario and starts to deviate from it only at redshift $z \sim 40$. This is because the CMBR-baryon interaction dominates over the DM-b interaction upto redshift $z \sim 40$. Consequently, the recombination rate is very similar to the standard scenario and starts to increase at redshifts $z \lesssim 40$. In this scenario the recombination rate becomes higher only for a shorter period of the cosmic time and, therefore, the deviation of the ionisation fraction from the standard scenario prediction is less. We, therefore, conclude that very different thermal histories of the IGM can be consistent with the EDGES 21-cm signal. However, the impact of them on the ionisation history could vary considerably. This prohibits us to accurately estimate the ionisation fraction during the cosmic dawn and dark ages.

Fig. \ref{fig:xfrac}b shows how ionisation fraction $x$ increases when X-ray photons from first luminous sources start ionizing the IGM. We see that $x$ starts increasing at redshift as early as $\sim 18$ for $f_{\star}=0.1$ whereas the visible impact of X-ray photons on $x$ is seen from redshift $\sim 14$ for $f_{\star}=0.01$. We discuss above (also in Fig. \ref{fig:T_21}) that the predicted IGM temperature and differential brightness temperature $T_b$ for $f_{\star} \gtrsim 0.1$ are inconsistent with the EDGES observations. The lower panel of Fig. \ref{fig:xfrac} plots the percentage difference of ionisation fractions between cases with and without the DM-baryon interaction in presence of X-ray heating. We show results only for $(m_{\chi},\sigma_{45})=(0.4 \, {\rm GeV}, 100)$. We find that the change is very similar to the case where X-ray heating is not included (see the bottom panel of Fig. \ref{fig:xfrac}a) for redshifts $z \gtrsim 22$. X-ray photons affects the ionisation fraction at redshifts lower than that. Early productions of X-ray photons is not allowed by the EDGES as they heat up the IGM much earlier which is inconsistent with EDGES profile. Late production of X-ray photons only change the ionisation fraction later. Impact of colder IGM on $x$ gets diluted at redshifts $z \lesssim 22$ because X-ray photons dominates the ionisation. 
 
 \begin{figure}[hbtp]
\includegraphics[width=8.5cm,angle=0]{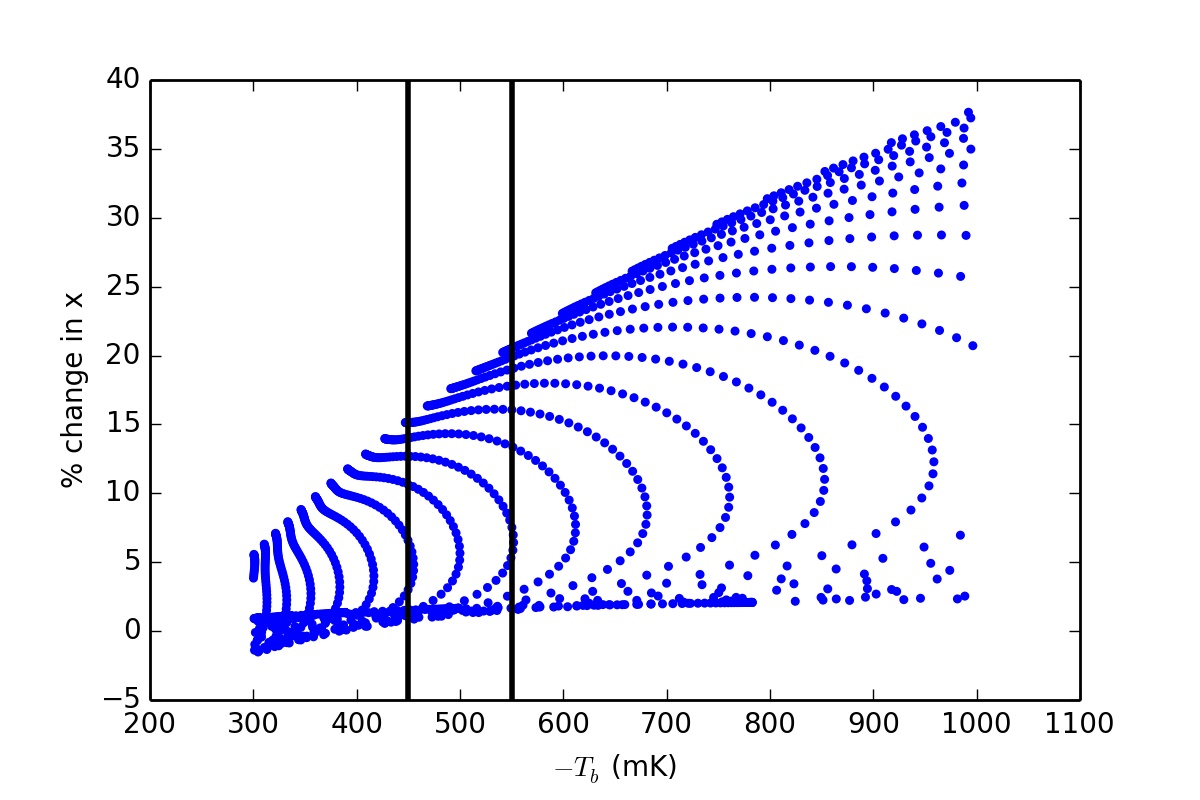}
 \caption{This shows the percentage change in the ionisation fraction w.r.t the standard scenario at redshift $z=17.2$ for all possible combinations of $(m_{\chi} \, , \sigma_{45})$ which predict the differential brightness temperature $T_b$ in the range between $-300$ mK and $-1000$ mK, allowed by the EDGES 21-cm profile. The two vertical lines  show the changes in the ionisation fraction $x$ if $T_b$ is restricted between  $-450$ mK and $-550$ mK. We find very similar results when effects due to X-ray heating with $f_{\star}=0.01$ is included.}
\label{fig:tb-xfrac}
\end{figure}

\begin{figure}[hbtp]

 \includegraphics[width=8.5cm,angle=0]{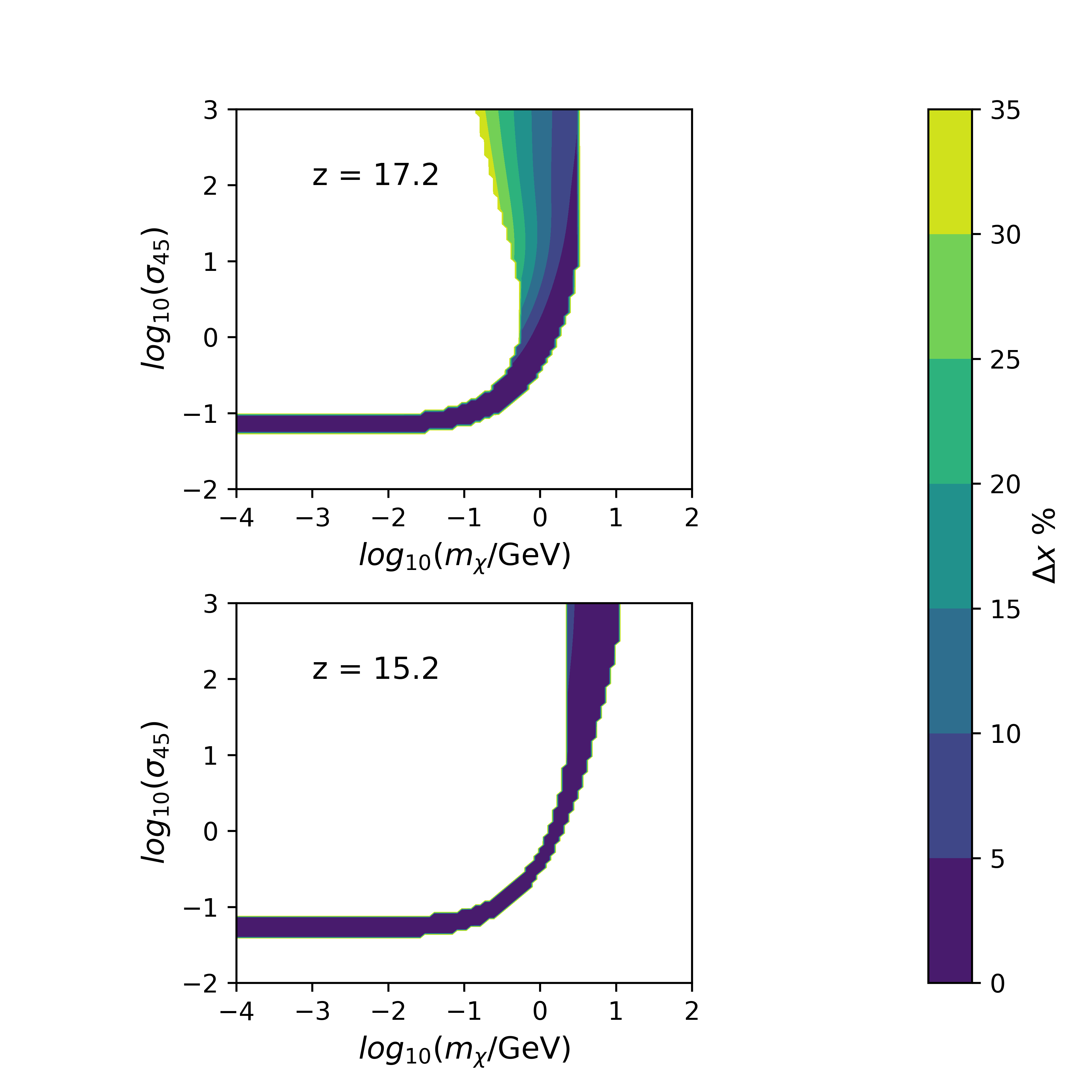}

 \caption{These contour plots show the percentage change in the ionisation fraction due to colder IGM w.r.t the standard scenario (with X-ray heating) at redshifts $z=17.2$ and $z=15.2$. In the upper panel, the white region is excluded by the  EDGES measurements at $99\%$ confidence level as the HI differential brightness temperature  $T_b$ is either lower than $-1000$ mK or higher than $-300$ mK in the region. The lower panel shows the allowed region for $T_b$ within $-191$ mK and $-371$ mK at $z=15.2$}
\label{fig:contour}
\end{figure}

\begin{figure}[hbtp]
    
    \includegraphics[width=8.5cm,angle=0]{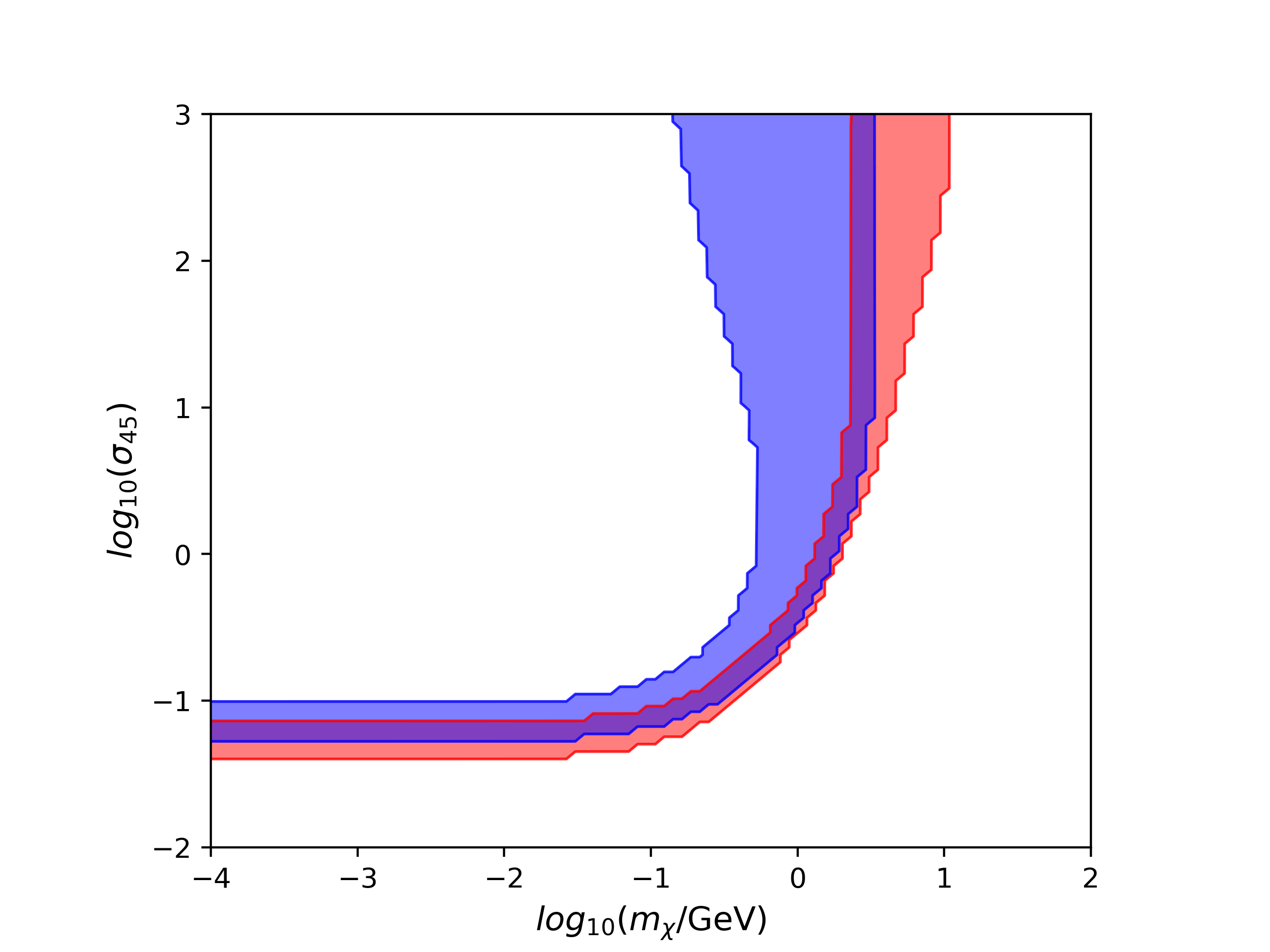}
    
    \caption{This plot shows the allowed regions of the parameter space at $z=17.2$(blue) and $z=15.2$(red). The narrow overlap region indicates the allowed region which explain EDGES measurements at both redshifts. The change in the ionisation fraction in the overlap region due to colder IGM is between $\sim 0$ to $\sim 5 \%$. We set the star formation efficiency $f_{\star}=0.01$ here.}
    
    \label{fig:Overlap}
\end{figure}

 We explore the entire $(m_{\chi}, \sigma_{45})$ parameter space for $f_{\star}$ = 0.01 and  present our results on the changes in $x$ in Fig. \ref{fig:tb-xfrac}, \ref{fig:contour}  and \ref{fig:Overlap}. Fig. \ref{fig:tb-xfrac} shows the percentage change in the ionisation fraction w.r.t the standard scenario at redshift $z=17.2$ for all possible combinations of $(m_{\chi} \, , \sigma_{45})$ which predict the differential brightness temperature $T_b$ in the range between $-300$ mK and $-1000$ mK which is allowed by the EDGES measurements. We see that the suppression in the ionisation fraction can be anything from $\sim 0 \%$ upto $\sim 36 \%$. We find very similar results if we switch off the X-ray heating. We see that large cross sections suppress the ionisation fraction more whereas smaller cross sections have very little impact on the ionisation fraction. This large uncertainty in estimating $x$ is reduced only by a small amount even if we restrict $T_b$ at $z=17.2$ between $-450$ mK and $-550$ mK. This suggests that precise estimation of the ionisation fraction seems difficult even for more accurate measurements of the global 21-cm signal at $z=17.2$. The uncertainty in estimating the ionisation fraction due to a wide range of possibilities of the IGM thermal histories also remains at higher redshifts. We, however, find that the uncertainty in estimating the change in $x$ may be reduced if we consider EDGES $T_b$ measurements at two or more redshifts simultaneously. Fig. \ref{fig:contour} shows the changes in the ionisation fraction in the $(m_{\chi} \,  \sigma_{45})$ parameter space at $z=17.2$ and $z=15.2$ for $f_{\star}=0.01$. The upper panel highlights the parameter space where the $T_b$ is within $-300$ mK and $-1000$ mK at redshift $z=17.2$. The lower panel corresponds the parameter space for  $-191 \, {\rm mK} > T_b > -317 \, {\rm mK}$  which is the allowed range measured by the EDGES at $z=15.2$. We use the half-maximum amplitude of $-254$ mK from the best-fit EDGES profile at $z=15.2$ and calculate the standard deviation of $63$ mK using the different hardware cases of the EDGES measurements. We notice that the change in $x$ due to colder IGM lies between $\sim 0 \%$ to $\sim 5 \%$, which is much lower if we use the EDGES measurements at $z=17.2$.

Fig. \ref{fig:Overlap} shows both the allowed regions in the parameter space and highlights the overlap between the two. The narrow overlap region in the parameter space is allowed by $T_b$ measurements at both redshifts.  We compare the region with that of Fig. \ref{fig:contour} and see that the change in $x$ due to colder IGM could be between $\sim 0 \%$ to $\sim 5 \%$ for the X-ray model we consider.  Thus the uncertainty in estimating $x$ is reduced substantially if we use measurements at multiple redshifts.  We can also put more stringent constraints on the parameter space by accurately measuring the global 21-cm signal history. However, we must mention that the results above are very specific to the X-ray heating model we consider. It might change for different models of heating. Thus a detailed study is required towards this. 

We further note in Fig. \ref{fig:temp} that the IGM temperatures $T_g$ at redshift $z \sim 50$ are quite different for the two scenarios although they predict similar $T_b$ during cosmic dawn. We anticipate that the IGM temperature measured at two different redshifts (e.g. during dark ages and cosmic dawn) can be used to put even more stringent constraints on the evolution of the IGM temperature and accurately estimate the ionisation fraction.

\section{Summary and Discussion}
Recent measurements of the global 21-cm signal from the cosmic dawn by the EDGES suggest that the IGM  can be significantly colder at redshift $z \sim 17$ compared to its expected value. The `colder IGM' scenario enhances the recombination (of neutral Hydrogen) rate and affects  the recombination history of the universe.  We study this, in detail, in the context of DM-b interaction model which is a promising way to explain the EDGES detection. 

We find that the hydrogen ionisation fraction gets suppressed for all possible combinations of the DM-b model parameters  ($m_{\chi}, \sigma_{45}$) which are consistent with EDGES measurements. Although the suppression is stronger during the cosmic dawn, we see that the effect is significant even at higher redshifts during the dark ages. However, the actual amount of suppression in the ionisation fraction during the cosmic dawn and dark ages depends on the entire thermal history of the IGM, from the epoch of thermal decoupling of hydrogen gas and  the CMBR to the cosmic dawn. It is possible that two scenarios which predict very similar HI differential brightness temperature at redshifts where the EDGES measured the 21-cm signal have completely different IGM and HI differential brightness temperature at higher redshifts. Consequently, the ionisation history is also different in these scenarios. We explore the entire parameter space $(m_{\chi}, \sigma_{45})$ of the DM-b interaction model and find that the suppression in the ionisation fraction at redshift $z \sim 17$,  w.r.t the standard scenario, i.e., without the DM-b interaction, could be anything between $\sim 0 \%$ to $\sim 36 \%$ for scenarios which are consistent with the EDGES measurements. We also see that this large uncertainty in estimating the ionisation fraction remains even for a more accurate measurement of 21-cm signal from the cosmic dawn.

The above result changes very little even if we include effects due to X-ray photons arising from first sources. The reason is that the recombination history depends on the entire temperature history of universe from the recombination epoch to the cosmic dawn. X-ray photons from first sources alters the temperature only at redshifts $z\lesssim 22$. Moreover, X-ray can't be efficient at early redshifts $\sim 22$ as it is disfavoured by the EDGES profile. Inclusion of X-ray heating allows us to compare the model predictions with the lower redshift half of the EDGES observed profile. This helps to put stronger constraints on the DM-baryon model parameters and  the ionisation fraction. We find that the model with $f_{\star} \sim 0.01$ is consistent with the rise of EDGES profile. Both lower and higher star formation efficiencies are ruled out since they are inconsistent with the rise of EDGES profile.

The suppressed ionisation fraction for a considerable duration of the recombination history has several implications. First, it affects the formation of molecules in the early universe \cite{2002lepp} and might have indirect influence on the early star formations. Second, the impact of the magnetic field on the early structure formation and universe's thermal history prior to the cosmic dawn could change. Further, the contribution of spatial fluctuations in the ionisation fraction to the total fluctuations in the HI 21-cm signal during the dark ages  \cite{2018ansar} is  likely to get affected. A thorough and detailed investigation is required to assess the impact of the suppressed ionisation fraction on all these important astrophysical observables. 

We would like to mention that results presented here, in principle, might change if we include other effects such as dark matter annihilation \cite{2018damico}, magnetic field \cite{2019bhatt, bera2020} along with the DM-b interaction, Ly-$\alpha$ processes or any other phenomena which can alter the IGM temperature. However, given the EDGES constraints on the HI differential brightness temperature and the fact that the DM-b model considered here brackets extreme cases, we do not expect the results to change considerably. Finally, there are also a few fundamentally different explanations for the  unusually strong signal such as the excess radio background \cite{2018feng, 2018pospelov}, axion induced cooling \cite{2018mohanty}. The ARCADE 2 \cite{fixsen2011} and LWA1 \cite{dowell2018} experiments have claimed to have detected an excess component of radio background at $3-90$ GHz and $40-80$ MHz respectively. Although the nature of this excess is still not certain \cite{subrahmanyan2013},  it could be due to emission from an extended halo-wind of our galaxy and many other similar galaxies. This excess radio signal can offer a solution to the strong EDGES signal \cite{mirocha2019}. It is unlikely that these alternative explanations  have significant effects on the thermal history of the IGM before the onset of first sources, and consequently on the recombination history.


\section*{Acknowledgements} We thank anonymous referees for their detailed and constructive comments which have helped us to improve the quality of the paper. We also thank Raghunath Ghara for his help in the estimation of effects due to X-ray photons. KKD acknowledges financial support from BRNS through a project grant (sanction no: 57/14/10/2019-
BRNS) and University Grants Commission (UGC), Govt. of India. 

\end{document}